\documentclass[letterpaper]{article} 
\usepackage{aaai25}  
\usepackage{times}  
\usepackage{helvet}  
\usepackage{courier}  
\usepackage[hyphens]{url}  
\usepackage{graphicx} 
\usepackage{natbib}  
\usepackage{caption} 
\usepackage{algorithm}
\usepackage{algorithmic}
\usepackage{booktabs}
\usepackage{amsmath}
\usepackage{amssymb}
\usepackage{tabularx}
\usepackage[table]{xcolor}
\usepackage{multirow} 

\usepackage{xcolor}

\usepackage{newfloat}
\usepackage{listings}
\DeclareCaptionStyle{ruled}{labelfont=normalfont,labelsep=colon,strut=off} 
\floatstyle{ruled}
\newfloat{listing}{tb}{lst}{}
\floatname{listing}{Listing}
\title{dyAb: Flow Matching for Flexible Antibody Design with AlphaFold-driven Pre-binding Antigen}
\author{
    Cheng Tan\equalcontrib \textsuperscript{\rm 1,2},
    Yijie Zhang\equalcontrib \textsuperscript{\rm 3,4},
    Zhangyang Gao\equalcontrib \textsuperscript{\rm 2},
    Yufei Huang\textsuperscript{\rm 2},
    Haitao Lin\textsuperscript{\rm 2},
    Lirong Wu\textsuperscript{\rm 2},
    Fandi Wu\textsuperscript{\rm 5},
    Mathieu Blanchette \textsuperscript{\rm 3,4},
    Stan Z. Li\thanks{Corresponding author}\textsuperscript{\rm 2}
}
\affiliations{
    \textsuperscript{\rm 1}Zhejiang University \\
    \textsuperscript{\rm 2}AI Lab, Westlake University \\ 
    \textsuperscript{\rm 3}School of Computer Science, McGill University \\
    \textsuperscript{\rm 4}MILA - Québec AI Institute \\
    \textsuperscript{\rm 5}Tencent, AI for Life Science Lab\\
    
}

\usepackage{bibentry}

\begin{document}

\maketitle

\begin{abstract}
The development of therapeutic antibodies heavily relies on accurate predictions of how antigens will interact with antibodies. Existing computational methods in antibody design often overlook crucial conformational changes that antigens undergo during the binding process, significantly impacting the reliability of the resulting antibodies. To bridge this gap, we introduce dyAb, a flexible framework that incorporates AlphaFold2-driven predictions to model pre-binding antigen structures and specifically addresses the dynamic nature of antigen conformation changes. Our dyAb model leverages a unique combination of coarse-grained interface alignment and fine-grained flow matching techniques to simulate the interaction dynamics and structural evolution of the antigen-antibody complex, providing a realistic representation of the binding process. Extensive experiments show that dyAb significantly outperforms existing models in antibody design involving changing antigen conformations. These results highlight dyAb's potential to streamline the design process for therapeutic antibodies, promising more efficient development cycles and improved outcomes in clinical applications.
\end{abstract}

%
\begin{links}
    \link{Code}{https://github.com/A4Bio/dyAb}
\end{links}

\section{Introduction}

Antibodies are pivotal components of the immune system, equipped to identify and neutralize foreign entities such as bacteria, viruses, and other pathogens~\cite{raybould2019five,kong2023end,shi2022protein}. These Y-shaped proteins possess two binding arms that latch onto antigens. Upon binding to an antigen, antibodies mark invaders for destruction by other immune cells, which is a pivotal process for the body's defense mechanism against infections~\cite{basu2019recombinant}. In therapeutic applications, the natural binding capability of antibodies is harnessed to develop targeted treatments for a myriad of diseases, ranging from various types of cancers to autoimmune disorders and infectious diseases~\cite{kuroda2012computer,tiller2015advances}. The fundamental role of antibodies in immune surveillance underscores their importance in vaccination, where exposure to a specific antigen primes the immune system for future encounters, offering protection against diseases~\cite{maynard2000antibody,akbar2022progress}.

Early efforts on antibody design primarily focused on generating sequences for the Complementarity-Determining Regions (CDRs) without modeling the corresponding three-dimensional structures~\cite{saka2021antibody, alley2019unified, shin2021protein}. RefineGNN~\cite{jiniterative} enables the co-design of both the sequences and structures of antibody CDRs. Further advancements were made with DiffAb~\cite{luoantigen}, which attempts to generate antibodies with high affinity to given antigen structures, and MEAN~\cite{kong2023conditional}, which incorporates light chain context as a conditional input to generate CDRs. Moreover, dyMEAN~\cite{kong2023end} further proposes an end-to-end full-atom antibody design model. Despite these developments, a notable flaw persists in these computational methodologies: \textit{they typically do not consider the dynamic alterations in antigen structures upon antibody binding}. Instead, they take the post-binding structures as the foundation for design, overlooking the inherent flexibility and conformational shifts that antigens undergo. This deficiency can lead to predictions that fail to accurately capture the dynamics of antibody-antigen interactions, potentially resulting in antibody designs that are less effective in real-world applications. In essence, existing methods are \textit{starting with the answer}, which limits their practical utility.

\begin{figure}
\centering
\includegraphics[width=0.44\textwidth]{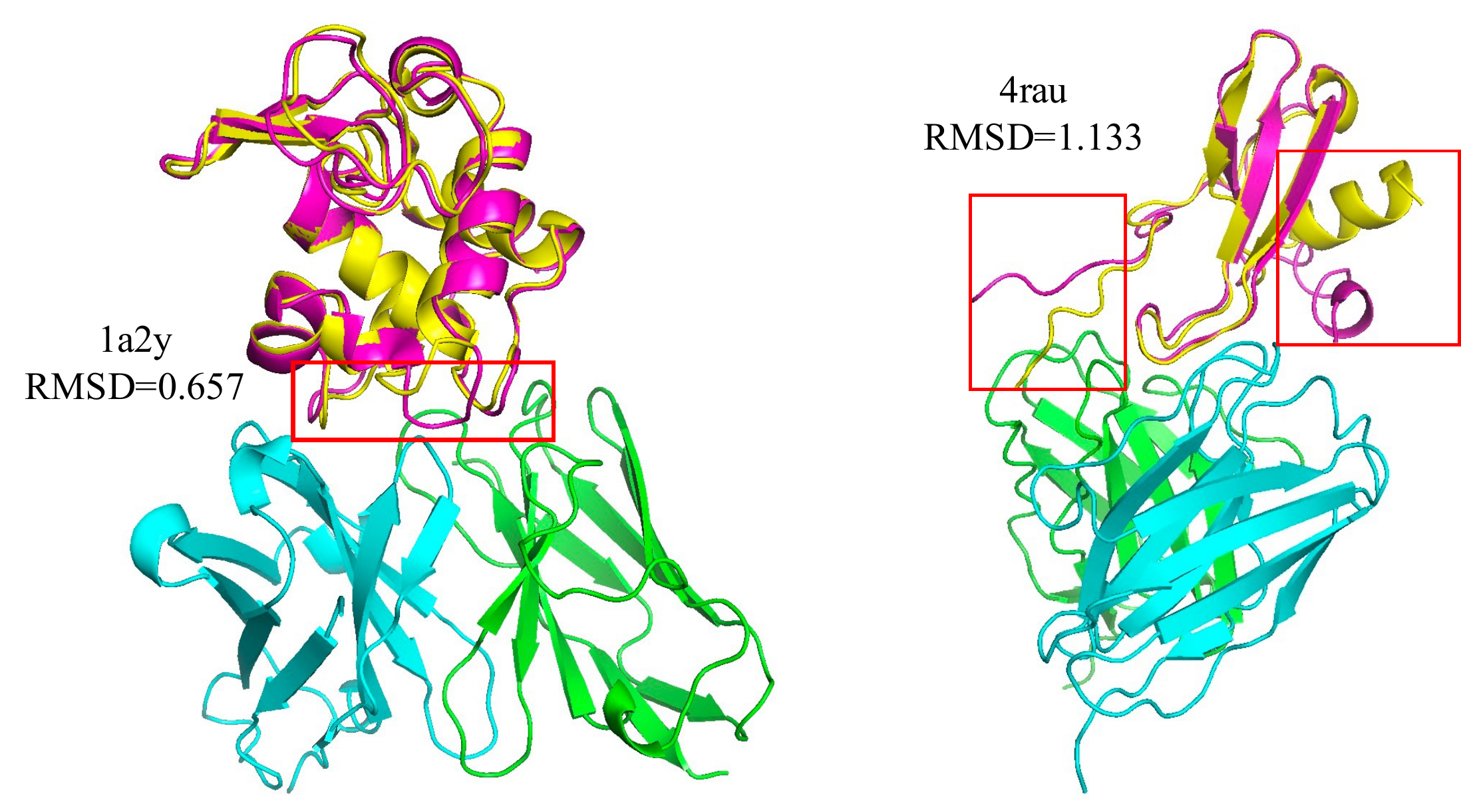}
\caption{Alignment of predicted and experimental antigen structures before and after binding. The pre-binding antigen structures, predicted by AlphaFold2, are depicted in yellow. The post-binding antigen structures, derived from experimental data of antigen-antibody complexes, are shown in red. The antibodies are colored in green and blue. The epitopes are highlighted with a red box.
}
\label{fig:motivation_example}
\end{figure}

The introduction of AlphaFold2~\cite{jumper2021highly,abramson2024accurate} marks a paradigm shift to predict protein structures with unprecedented accuracy and efficiency. AlphaFold2's predictions have been widely adopted in various computational biology applications, including protein folding, protein-protein interactions, and protein design~\cite{varadi2022alphafold,hu2022protein,hsu2022learning}. Its high-fidelity structural predictions offer a valuable starting point for modeling the pre-binding conformations of antigens. By employing the predicted antigen structures from AlphaFold2, we can ensure the reliability of the initial antigen conformation. Figure~\ref{fig:motivation_example} shows that while the root-mean-square deviation (RMSD) between AlphaFold2-predicted structures and experimental post-binding structures of antigens is minimal, significant discrepancies often occur in the interface regions where antibodies bind. This observation underscores the critical need to account for the dynamic nature of antigen structures during antibody design.

We introduce dyAb, which leverages AlphaFold2's predictions to model pre-binding antigen structures and explicitly addresses the dynamic nature of antigen conformation changes when designing the antibody. dyAb combines a coarse-grained interface alignment with a fine-grained flow matching approach to simulate the interaction dynamics and structural evolution of the antigen-antibody complex, providing a more realistic representation of the binding process. Extensive experiments demonstrate that dyAb significantly outperforms existing models in antibody design when involving changing antigen conformations, promising more efficient development cycles in applications.

\section{Related Work}

\paragraph{Protein Design} 

Several approaches in structure-based protein design leverage fragment-based and energy-based features derived from protein structures~\cite{wang2021directed,hu2022protein}. StructGNN~\cite{ingraham2019generative} introduced a paradigm shift by framing fixed-backbone design as a structure-to-sequence problem. GVP~\cite{jinglearning} introduced architectures with translational and rotational equivariances. GCA~\cite{tan2022generative} utilized global attention to learn geometric representations from residue interactions. AlphaDesign~\cite{gao2022alphadesign} established a protein design benchmark based on AlphaFold DB~\cite{varadi2022alphafold,jumper2021highly}. ESM-IF~\cite{hsu2022learning} augmented training data by incorporating predicted structures from AlphaFold2~\cite{jumper2021highly}. ProteinMPNN~\cite{dauparas2022robust} employed expressive structural features with message-passing neural networks. PiFold~\cite{gao2023pifold} introduced additional structural features and generated protein sequences in one shot. Our focus is on antibody design, a specialized area within protein design.
\paragraph{Antibody Design}


Early approaches used Monte Carlo simulations to iteratively update sequences and structures~\cite{pantazes2010optcdr,adolf2018rosettaantibodydesign,warszawski2019optimizing,ruffolo2021deciphering}, but these were computationally expensive and prone to local energy minima. Deep generative models have since emerged as viable alternatives. Sequence-based methods~\cite{alley2019unified,saka2021antibody,shin2021protein,akbar2022silico} paved the way, followed by more advanced techniques like RefineGNN~\cite{jiniterative} for CDR co-design, DiffAb~\cite{luoantigen} for antigen-specific antibody generation using diffusion models, and CEM~\cite{fu2022antibody} for modeling CDR geometry constraints. MEAN~\cite{kong2023conditional} and dyMEAN~\cite{kong2023end} introduced E(3)-equivariant message passing and full-atom design models, respectively. tFold~\cite{wu2022tfold} leverages protein language models for antibody-antigen structure prediction, and \citet{kim2024anfinsen} proposed a decoupled sequence-structure generation method. However, these methods do not explicitly account for dynamic antigen structural changes upon antibody binding.

\paragraph{AlphaFold2 Benefits Computational Biology} 

Using AlphaFold2's predicted structures has become a common practice in various computational biology applications. ESM-IF~\cite{hsu2022learning} augment training data by predicting structures using AlphaFold2 for protein sequence design. By subsampling the multiple sequence alignments (MSA) input to AlphaFold2, approaches like those discussed in~\citet{del2022sampling} have successfully predicted alternative conformations. AF-Cluster~\cite{wayment2024predicting} applies sequence similarity clustering to predict alternative states of metamorphic proteins. AlphaFlow combines the structural prediction power of AlphaFold2 with flow matching techniques to generate protein ensembles~\cite{jing2024alphafold}.



\section{Background}

\subsection{Antibody-antigen Complex}

Proteins are biological molecules consisting of one or more chains of amino acid residues. These residues are the basic building blocks of proteins, each represented by one of 20 standard amino acids. In this context, a protein complex can be described as comprising $N$ amino acids, each denoted as a residue, and collectively forming a sequence $\mathcal{S} = \{s_i\}_{i=1}^N$. The three-dimensional structure of a protein is captured through the coordinates of its backbone atoms, specifically denoted as $\mathcal{X} = \{ \boldsymbol{x}_{i}\}_{i=1}^{N}$, where $\boldsymbol{x}_{i}\in\mathbb{R}^{3 \times c_i}$ and $c_i$ represents the number of atoms in the $i$-th residue. An antibody-antigen complex, a typical type of protein complex, can be defined by the pair $\mathcal{C} = (\mathcal{S}, \mathcal{X})$.

Within this complex, the antibody and the antigen play distinct roles: (i) Antibody, $\mathcal{C}_{ab} = (\mathcal{S}_{ab}, \mathcal{X}_{ab})$, is Y-shaped symmetric protein composed of two identical sets of chains, each set containing a heavy (H) chain and a light (L) chain. These chains are further divided into several constant domains and a variable domain. The variable domains denoted as $V_H$ and $V_L$ for the heavy and light chains respectively, include regions known as framework regions (FRs) and complementarity-determining regions (CDRs). (ii) Antigen, $\mathcal{C}_{ag} = (\mathcal{S}_{ag}, \mathcal{X}_{ag})$, is a protein that when bound by an antibody, forms a complex that can elicit an immune response. The antibody-antigen complex, $\mathcal{C} = \mathcal{C}_{ab} \cup \mathcal{C}_{ag}$, results from the interaction between the antibody and the antigen.

\subsection{Problem Statement}

In this work, we focus on the challenge of designing antibodies that not only bind to antigens with high affinity but also incorporate the dynamic nature of antigen structures during the binding process. Traditional approaches typically focus on static models of antigen structures, often neglecting the significant conformational changes that antigens undergo upon interaction with antibodies:
\begin{equation}
  \mathcal{F}: \mathcal{C}_{ag} \rightarrow \mathcal{C}_{ab} \cup \mathcal{C}_{ag},
\end{equation}
These approaches use the post-binding antigen structure as the basis for antibody design. Recognizing this gap, we propose a redefined problem statement that considers both the pre-binding and post-binding states of the antigen structures:
\begin{equation}
  \mathcal{F}: \mathcal{C}_{ag}^{(0)} \rightarrow \mathcal{C}_{ab} \cup \mathcal{C}_{ag}^{(1)},
\end{equation}
Here, $\mathcal{C}_{ag}^{(0)}$ explicitly represents the pre-binding antigen structure, and $\mathcal{C}_{ag}^{(1)}$ denotes the antigen structure after antibody binding. In practice, we focus on the epitope of the antigen and the variable domains of the antibody and model them as graphs $\mathcal{G}_E(\mathcal{V}_E,\mathcal{E}_E)$ and $\mathcal{G}_A(\mathcal{V}_A, \mathcal{E}_A)$, respectively. Here, $\mathcal{V}_E$ and $\mathcal{V}_A$ capture intra-residue level features, while $\mathcal{E}_E$ and $\mathcal{E}_A$ represent inter-residue level features. They are derived from the sequence and structural data $\mathcal{S}$ and $\mathcal{X}$. The connectivity between residues is built using the k-nearest neighbors (kNN) approach, which calculates the minimum pairwise distance between all atoms in residues $i$ and $j$:
\begin{equation}
\begin{aligned}
d(v_i, v_j) = \min_{1\leq p \leq c_i, 1 \leq q \leq c_j} \|\mathcal{X}_{i}(:,p) - \mathcal{X}_{j}(:,q)\|_2,
\end{aligned}
\end{equation}
where $\mathcal{X}_{i}(:,p)$ is the coordinates of the $p$-th atom in $\mathcal{X}_i$.

\subsection{Flow Matching}
\label{bg:flow_matching}
Flow matching~\cite{lipman2022flow,albergo2023stochastic,liu2022flow} is a generative modeling paradigm that has been inspired by and further extends the notable successes of diffusion models in image~\cite{ho2020denoising,song2021score} and molecule~\cite{jing2024alphafold,stark2023harmonic,lin2024ppflow} domains. This technique is grounded in the fundamental objective of learning an ordinary differential equation (ODE) to transform a prior distribution $p_0$ into a target data distribution $p_1$ over a defined time interval from $t=0$ to $t=1$. Let $\mathcal{P}$ denote the space of probability functions over a manifold $\mathcal{M}$ equipped with a Riemannian metric $g$. The transformation is a probability path $p_t: [0, 1] \rightarrow \mathcal{P}$ on $\mathcal{M}$, interpolating between $p_0$ and $p_1$. At any time $t$, the corresponding gradient vector $u_t(\boldsymbol{x})$ at a point $\boldsymbol{x}$ in $\mathcal{M}$ lies in the tangent space $\mathcal{T}_{\boldsymbol{x}}\mathcal{M}$. 

To approximate this vector field, a flow matching tangent vector field $v_t: [0,1] \times \mathcal{M} \rightarrow \mathcal{M}$ is employed, parameterized by $\theta$. The objective is to minimize the loss function $\mathcal{L}_{RFM}(\theta) = \mathbb{E}_{t, p_t(\boldsymbol{x})}\|v_t(\boldsymbol{x}) - u_t(\boldsymbol{x})\|_g^2$, which quantifies the discrepancy between the learned vector field and the true gradient vectors. Given the intractable nature of $u_t(\boldsymbol x)$, flow matching leverages a conditional density path $p_t(\boldsymbol{x} | \boldsymbol{x}_1)$ and employs a conditional flow matching objective:
\begin{equation}
\mathcal{L}_{CRFM} = \mathbb{E}_{t\sim\mathcal{U}(0,1), p_1(\boldsymbol{x}_1), p_t(\boldsymbol{x}|\boldsymbol{x}_1)} \|v_t(\boldsymbol{x}) - u_t(\boldsymbol{x}|\boldsymbol{x}_0,\boldsymbol{x}_1)\|_g^2.
\label{fm_loss}
\end{equation}

\section{dyAb}

\begin{figure}[ht]
\centering
\includegraphics[width=1.0\linewidth]{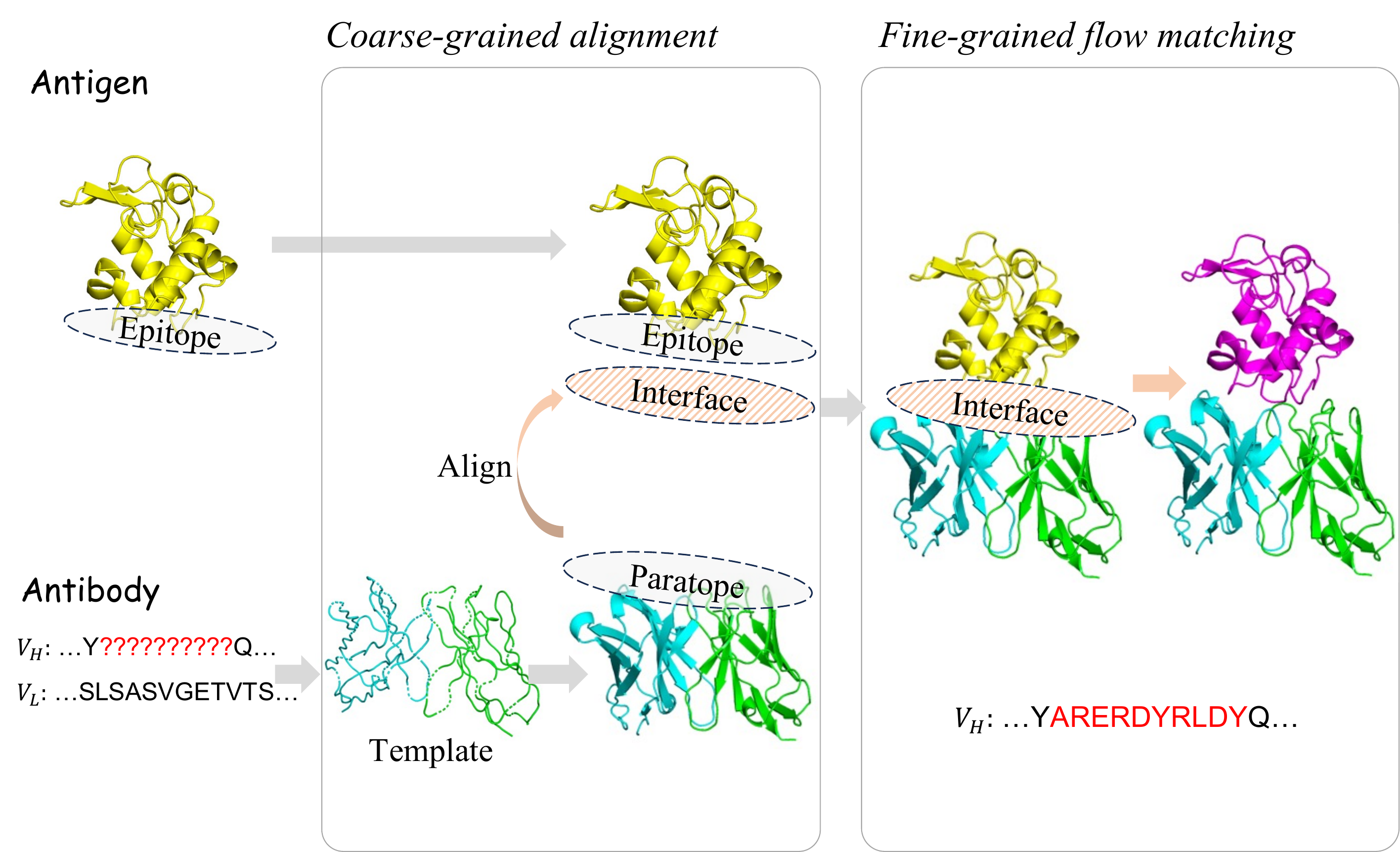}
\caption{The overview framework of dyAb. The pre-binding antigen structures are predicted by AlphaFold2 and used as input to the model. dyAb consists of two main components: coarse-grained interface alignment and fine-grained flow matching. The model is trained end-to-end to predict the post-binding antibody-antigen structures and the designed antibody sequences.}
\label{fig:overview}
\end{figure}

In this work, we introduce dyAb, a framework designed to address the dynamic nature of antigen-antibody interactions. Our approach integrates AlphaFold2-driven pre-binding antigen structures with a unique combination of coarse-grained interface alignment and fine-grained sequence-structure flow matching techniques. This section provides an overview of the proposed dyAb framework and its components, as illustrated in Figure~\ref{fig:overview}.

The fundamental idea behind dyAb stems from the understanding that the nature of antigen-antibody interactions varies significantly depending on the spatial proximity between the two molecules. When an antibody and an antigen are far apart, their interactions are primarily influenced by macroscopic forces such as electrostatic attraction and hydrophobic effects. These forces are relatively simple and do not significantly alter the internal structures of the antigen and antibody. However, as the antibody and antigen come into closer proximity, the complexity of their interactions increases dramatically. The binding process becomes dominated by fine, atomic-level forces, including hydrogen bonding, van der Waals forces, and precise steric fits. These interactions necessitate a detailed and accurate modeling approach to capture the dynamic conformational changes and ensure high-affinity binding.

Recognizing this transition from macroscopic to microscopic interactions, dyAb employs a two-stage strategy: (i) coarse-grained interface alignment, which focuses on the macroscopic interactions, and (ii) fine-grained sequence-structure flow matching, which captures the microscopic interactions and structural evolution. dyAb provides a comprehensive representation of the antigen-antibody binding process, enabling the design of high-affinity antibodies that effectively target antigens.

\begin{figure*}[ht]
  \centering
  \includegraphics[width=\textwidth]{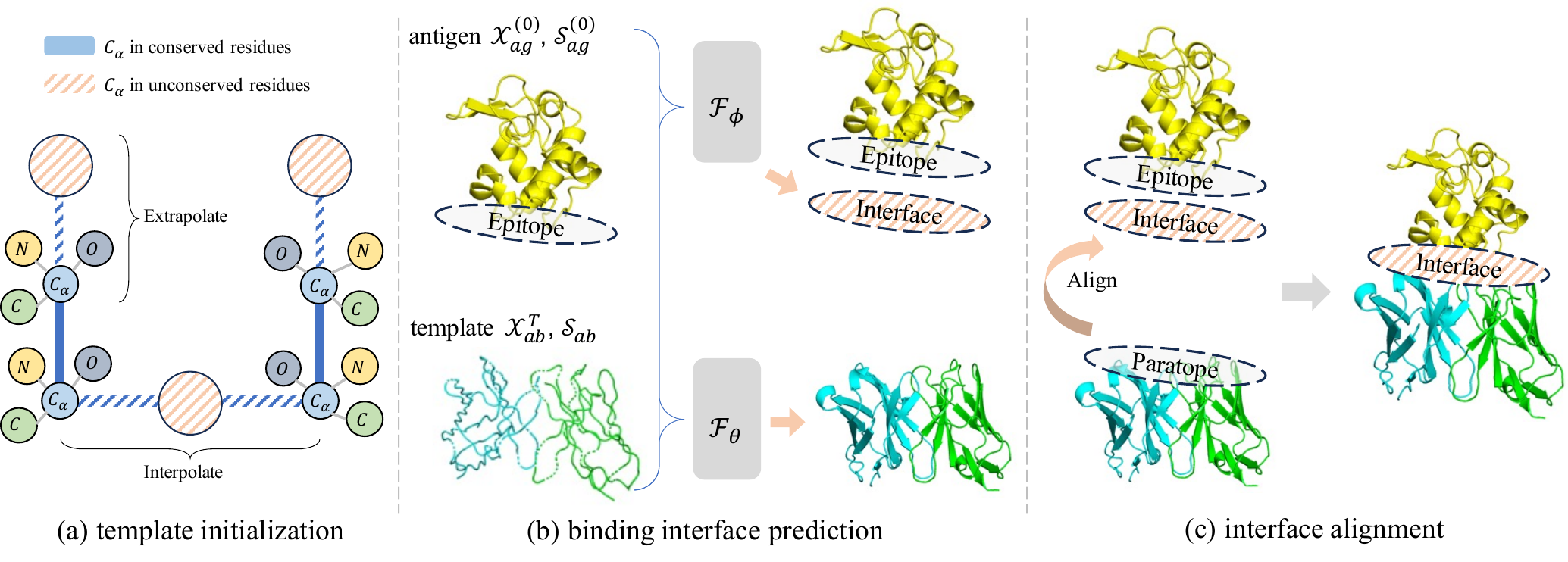}
  \caption{The key steps of the coarse-grained interface alignment process. (a) The pre-binding antigen structure is used to initialize the antibody structure. (b) The antibody structure and the binding interface are predicted by individual models. (c) The antibody structure is aligned to the predicted interface to generate the coarse-grained antibody-antigen complex.}
  \label{fig:coarse}
  \end{figure*}  

\subsection{Coarse-grained Interface Alignment}

In the initial stage of coarse-grained interface alignment, the objective is to establish an initial binding complex for the antibody and antigen based on the given pre-binding antigen structure and antibody sequences. At this stage, the antibody and antigen are relatively far apart, and their interactions are governed by macroscopic forces such as electrostatic attraction and hydrophobic effects. We assume that these interactions are simple and do not significantly alter their internal structures. The key steps of the coarse-grained interface alignment process are illustrated in Figure~\ref{fig:coarse}.

\paragraph{Antibody Structure Template Initialization} Following previous works~\cite{luoantigen,kong2023end,kong2023conditional,tan2024cross}, we begin with the pre-binding antigen $\mathcal{C}_{ag}^{(0)} = (\mathcal{S}_{ag}^{(0)}, \mathcal{X}_{ag}^{(0)})$. To initialize the antibody structure, we leverage the well-conserved framework regions by utilizing an average backbone template $\mathcal{X}^T_{ab}$ based on the IMGT numbering system~\cite{lefranc2003imgt}, which accurately identifies and aligns conserved residues. Conserved residues, crucial for maintaining the structural integrity of the antibody, are directly set according to the positions in the average backbone template $\mathcal{X}^T_{ab}$. For the remaining residues, a residue at position $k$ (either between or outside the conserved residues), its position $\boldsymbol{x}_k$ is determined by:
\begin{equation}
  \boldsymbol{x}_k = \boldsymbol{x}_{i} + \frac{k-i}{j-i}(\boldsymbol{x}_{j} - \boldsymbol{x}_{i}), \forall \boldsymbol{x}_k \in \mathcal{X}_{ab}^T,
\end{equation}
where $\boldsymbol{x}_{i}$ and $\boldsymbol{x}_{j}$ are the positions of two nearest conserved residues $i$ and $j$. For residues situated between the two nearest conserved residues ($i < k < j$), the position $\boldsymbol{x}_k$ is linearly interpolated between $\boldsymbol{x}_{i}$ and $\boldsymbol{x}_{j}$. For residues located at the termini of the antibody chains ($k < i$ or $k > j$), the position $\boldsymbol{x}_k$ is extrapolated based on the same interval defined by the nearest conserved residues. The coordinates of the backbone atoms in these residues are initialized by the coordinates of C$\alpha$ atoms.

\paragraph{Binding Interface Prediction} We employ individual models to predict the full-atom geometry of the antibody and the interface between the antibody and the antigen, which can be formulated as:
\begin{equation}
\begin{split}
\mathcal{S}_{ab}', \mathcal{X}_{ab}' &= \mathcal{F}_\theta (\mathcal{S}_{ab}, \mathcal{X}_{ab}^T, \mathcal{S}_{ag}^{(0)}, \mathcal{X}_{ag}^{(0)}), \\
\mathcal{S}_{itf}', \mathcal{X}_{itf}' &= \mathcal{F}_\phi (\mathcal{S}_{ab}, \mathcal{X}_{ab}^T, \mathcal{S}_{ag}^{(0)}, \mathcal{X}_{ag}^{(0)}),
\end{split}
\end{equation}

where $\mathcal{F}_\theta$ and $\mathcal{F}_\phi$ are the models used to predict the antibody structure and the binding interface. While the interface is expected to correspond to the paratope of the antibody, a key distinction is that $\mathcal{X}_{ab}'$ is built around the template $\mathcal{X}^T_{ab}$ whereas $\mathcal{X}_{itf}'$ is built around the epitope of the antigen $\mathcal{X}^{(0)}_{ag}$.

\paragraph{Interface Alignment} To align the antibody to the predicted interface, we utilize the root-mean-square deviation (RMSD) alignment approach~\cite{kabsch1976solution}. This involves using the Kabsch algorithm to find the optimal rotation and translation that minimize the RMSD between the two sets of points:
\begin{equation}
\begin{aligned}
\boldsymbol{Q}, \boldsymbol{t} &= \mathrm{Kabsch}(\mathcal{X}_{itf}', \mathcal{X}_{ab}'),\\
\mathcal{X}_{ab}^{(0)} &= \boldsymbol{Q}\mathcal{X}_{ab}' + \boldsymbol{t},
\end{aligned}
\end{equation}
where $\boldsymbol{Q} \in \mathbb{R}^{3\times 3}$ represents the optimal rotation matrix, and $\boldsymbol{t}\in\mathbb{R}^{3\times 1}$ is the translation vector. This alignment ensures that the antibody's paratope is correctly positioned relative to the antigen's epitope.

\subsection{Fine-grained Sequence-Structure Flow Matching}

\begin{figure}[ht]
\centering
\includegraphics[width=1.0\linewidth]{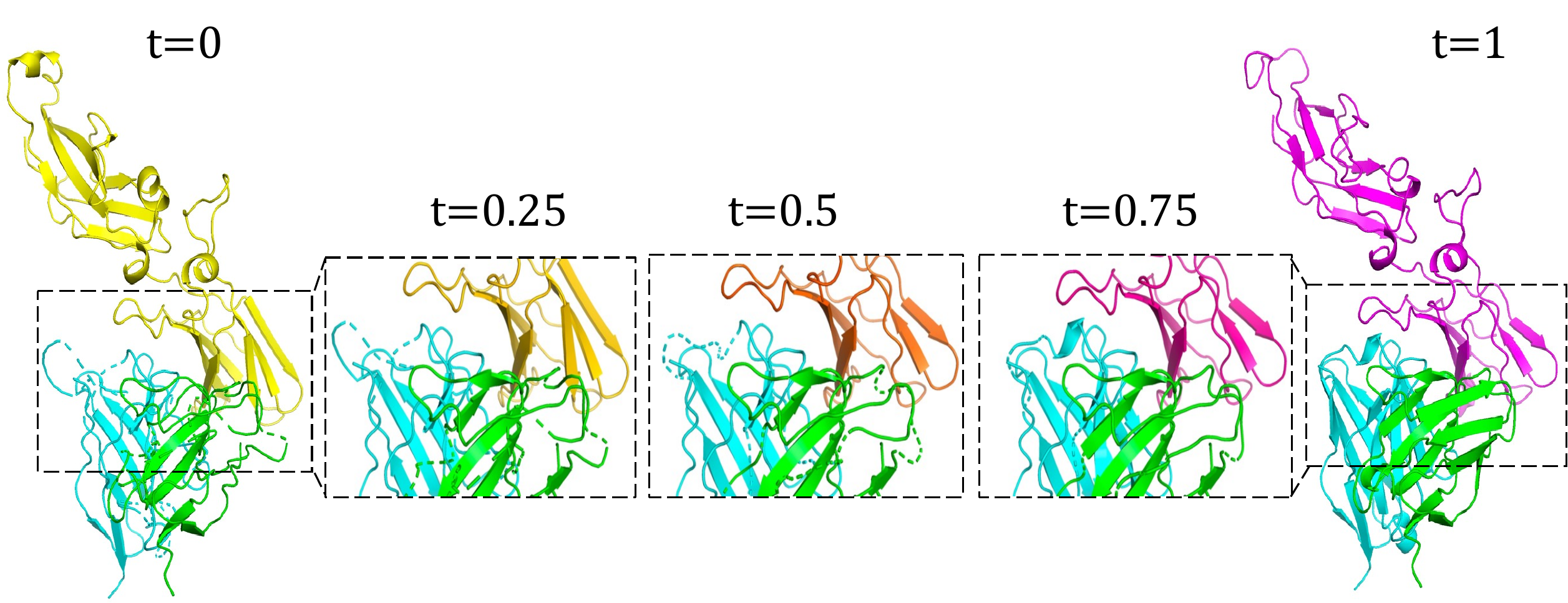}
\caption{Fine-grained iterative refinement process of the antibody-antigen complex. The interfaces of both the antigen and the antibody are iteratively refined.}
\label{fig:fine}
\end{figure}

The fine-grained flow matching stage is designed to capture the detailed atomic-level interactions and dynamic conformational changes that occur during antigen-antibody binding. As the antigen and antibody come into close proximity, their interactions become dominated by precise atomic-level forces. Given this proximity and the refined initial alignment, we employ an Euler method-based ODE solver to simulate the ultimate binding state in Euclidean space~\cite{das1994design,jing2024alphafold,stark2023harmonic}. The initial coarse-grained aligned structure serves as a close approximation to the ground truth, enabling the use of direct end-state prediction rather than iteratively predicting the transformation vector field. We show the iterative process of the structure evolution in Figure~\ref{fig:fine}.

\paragraph{Flow Matching on Structure}

In this approach, instead of predicting the vector field $u_t(\boldsymbol{x}|\boldsymbol{x}_0,\boldsymbol{x}_1)$, we directly predict the end state $\boldsymbol{x}_1$ of the binding process. The loss function is then modified to:
\begin{equation}
  \mathcal{L}_{Str} = \mathbb{E}_{t\sim\mathcal{U}(0,1), p_0\sim(\boldsymbol{x}_0), p_1\sim(\boldsymbol{x}_1), p_t(\boldsymbol{x}|\boldsymbol{x}_0, \boldsymbol{x}_1)} \|v^{str}_t(\boldsymbol{x}) - \boldsymbol{x}_1\|_2^2.
\end{equation}
The state update mechanism, analogous to the Euler method, is given by:
\begin{equation}
  \boldsymbol{x}_{t+\Delta t} = \boldsymbol{x}_t + \Delta t(\boldsymbol{\hat{x}}_1 - \boldsymbol{x}_0),
\end{equation}
where $\boldsymbol{\hat{x}}_1 = v^{str}_t(\boldsymbol{x})$ is  the predicted end state at time $t$. It ensures a smooth and controlled evolution of the state towards the predicted end state, leveraging the stability and simplicity of the Euler method. The detailed analysis of this flow matching approach is presented in Appendix.

\paragraph{Flow Matching on Sequence} 
Similarly, for the amino acid sequence of the antibody CDR regions $s \in \mathcal{S}_{ab}$, we directly model the probability vector of each residue type at each position. Here, $\boldsymbol{c}_t$ represents the probability vector of a multinomial distribution where $s_t \sim p(\boldsymbol{c}_t)$. We set $\boldsymbol{c}_1 = \mathrm{onehot}(s_i)$ and $\boldsymbol{c}_0 = (\frac{1}{20}, ..., \frac{1}{20})$. The loss function for this sequence flow matching is defined as:
\begin{equation}
  \mathcal{L}_{Seq} = \mathbb{E}_{t\sim\mathcal{U}(0,1), p_0\sim(\boldsymbol{c}_0), p_1\sim(\boldsymbol{c}_1), p_t(\boldsymbol{c}|\boldsymbol{c}_0, \boldsymbol{c}_1)} \|v^{seq}_t(\boldsymbol{x}) - \boldsymbol{c}_1\|_2^2.
\end{equation}
The sequence probabilities are updated iteratively, refining the predicted sequence $c_t$ at each step to closely match the desired distribution $\boldsymbol{c}_1$:
\begin{equation}
  \boldsymbol{c}_{t+\Delta t} = \boldsymbol{c}_t + \Delta t(\boldsymbol{\hat{c}}_1 - \boldsymbol{c}_0),
\end{equation}
where $\boldsymbol{\hat{c}}_1 = v^{seq}_t(\boldsymbol{c})$ is the predicted end state at time $t$. This iterative refinement process ensures that the sequence evolves towards the high-affinity state represented by $\boldsymbol{c}_1$.

\subsection{Model and Loss Function}

\begin{figure}[h]
  \centering
  \includegraphics[width=1.02\linewidth]{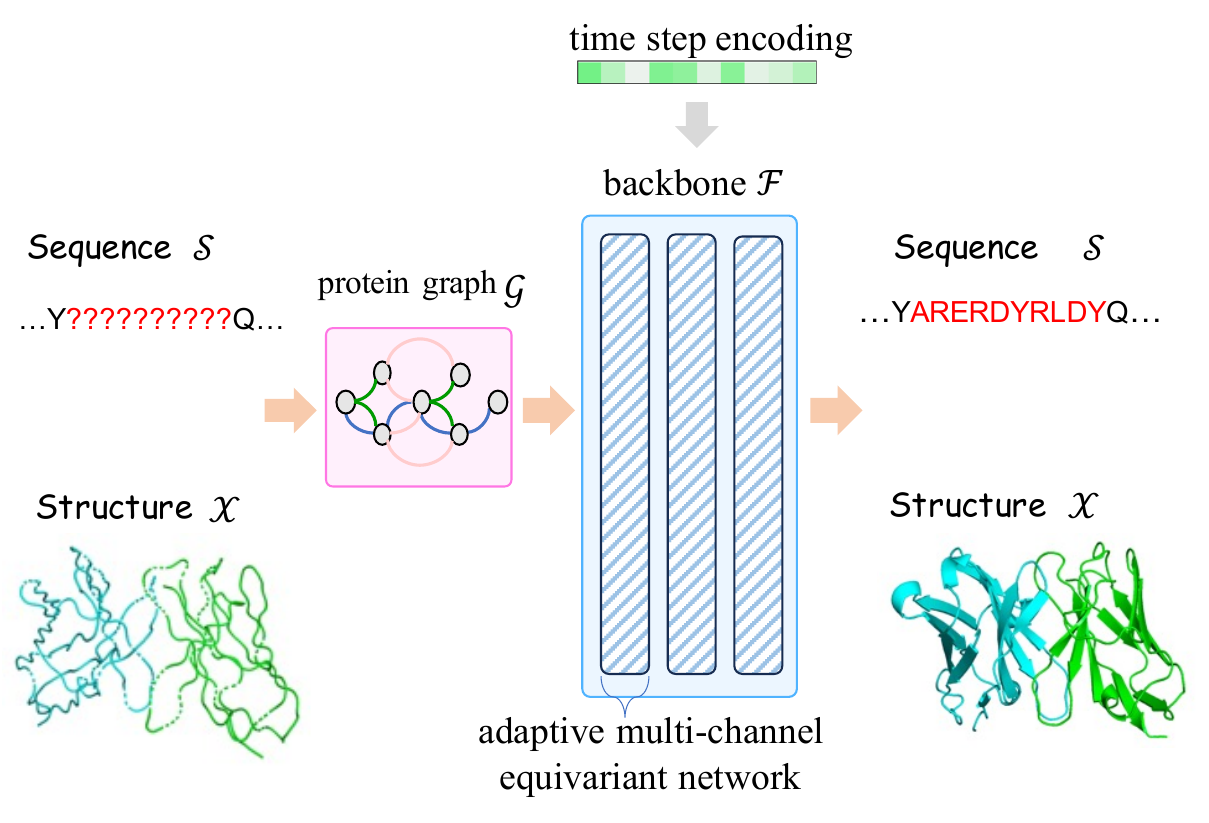}
  \caption{The overall model architecture of dyAb.}
  \label{fig:model}
\end{figure}

As shown in Figure~\ref{fig:model}, we employ the adaptive multi-channel equivariant network as the backbone network~\cite{kong2023conditional,kong2023end,han2024hemenet,kong2024full} that inputs the sequence and structure coordinates, modeling them as graphs $\mathcal{G}_E$ (epitope), $\mathcal{G}_A$ (antibody), and outputs the predicted sequence and structure. This network is designed to handle the rotational and translational symmetries inherent in molecular structures.

The total loss function comprises several components aimed at optimizing different aspects of the model. These include sequence and structure flow matching losses as well as interface alignment losses. The interface alignment loss $\mathcal{L}_{ITF} = \mathcal{L}_{sp} + \mathcal{L}_{dist}$ ensures that the predicted interface between the antibody and antigen is accurate. This loss has two subcomponents: (i) coordinate loss ($\mathcal{L}_{sp}$), Measures the difference between predicted and actual interface coordinates using the Huber loss~\cite{huber1992robust}:
\begin{equation}
  \mathcal{L}_{sp} = \frac{1}{|\mathcal{X}_{itf}|}\sum_{i \in \mathcal{X}_{itf}} \ell_{\text{huber}}(\boldsymbol{x}_i, \boldsymbol{x}_i^{*}),
\end{equation}
where $\mathcal{X}_{itf}$ are the coordinates of the predicted interface and $\boldsymbol{x}_i^{*}$ are the ground truth interface coordinates.
(ii) distance loss ($\mathcal{L}_{dist}$): Measures the distance between the predicted interface and the epitope of the antigen using the Huber loss:
\begin{equation}
  \mathcal{L}_{dist} = \frac{1}{|\mathcal{X}_{itf}| |\mathcal{X}_{ep}|} \sum_{i \in \mathcal{X}_{itf}, j \in \mathcal{X}{ep}} \ell_{\text{huber}}(d(i,j), d^*(i,j)),
\end{equation}
where $d(i, j)$ is the predicted distance between atoms $i$ and $j$, and $d^*(i, j)$ is the ground truth distance. The epitope coordinates are denoted by $\mathcal{X}_{ep}$.

The overall loss function $\mathcal{L}$ is a linear combination of the sequence flow matching loss $\mathcal{L}_{Seq}$, structure flow matching loss $\mathcal{L}_{Str}$, and interface alignment loss $\mathcal{L}_{ITF}$:
\begin{equation}
  \mathcal{L}_{total} = \mathcal{L}_{Seq} + \mathcal{L}_{Str} + \mathcal{L}_{ITF},
\end{equation}
This combined loss function ensures that the model accurately predicts both the sequence and structure of the antibody while maintaining a realistic and high-affinity binding interface with the antigen.

\section{Experiments}
\label{sec:exp}

We conduct comprehensive experiments that take into account the dynamic nature of antigen conformational changes. This novel experimental setting aims to emulate the flexible conditions under which antibodies must function. While the dataset split is consistent with previous works~\cite{kong2023end} for a fair comparison, we replace the pre-binding antigen structures with AlphaFold2's predictions and predict the ground-truth antigen-antibody complexes. We assess the models' capabilities in antibody design through three critical tasks: (i) Epitope-binding CDR-H3 generation, assessing the ability to generate the highly variable and functionally crucial CDR-H3 region. (2) Affinity optimization, evaluating the improvements in binding affinity of the designed antibodies to target antigens. (3) Complex structure prediction, predicting the 3D structure of the antibody-antigen complex. Moreover, a detailed ablation study of our proposed dyAb model is conducted to validate its effectiveness.

Our primary focus is on comparing the performance of our proposed dyAb model with dyMEAN~\cite{kong2023end}, as both are end-to-end models designed to streamline the antibody design process. Additionally, we also benchmark dyAb against other CDR generation baselines that require a multi-stage pipeline, including antibody structure prediction with IgFold~\cite{ruffolo2021deciphering}, antibody-antigen docking with HDock~\cite{yan2017hdock}, antibody CDR generation, and side-chain packing with Rosetta~\cite{das2008macromolecular}. The CDR generation baselines include (i) RosettaAb~\cite{adolf2018rosettaantibodydesign} is a traditional computational approach tailored for antibody design. (ii) HERN~\cite{jin2022antibody} employs a hierarchical message passing to predict atomic forces and use them to refine a binding complex in an iterative, equivariant manner. (iii) DiffAb~\cite{luoantigen} uses a diffusion model to generate antibodies targeting specific antigen structures. (iv) MEAN~\cite{kong2023conditional} employs E(3)-equivariant message passing and attention mechanisms to generate antibodies. 

\subsection{Epitope-binding CDR-H3 Generation}

CDR-H3 plays a pivotal role in determining the binding affinity of antibodies to their target antigens. Due to its high variability and central role in antigen binding, accurately generating its sequence and structure is a fundamental task in antibody design. We train the models on the Structural Antibody Database (SAbDab)~\cite{dunbar2014sabdab} and evaluate them on the RABD Benchmark~\cite{adolf2018rosettaantibodydesign}, which contains 60 antibody-antigen complexes annotated by biological experts. The input to the model is the pre-binding antigen structure predicted by AlphaFold2 and the incomplete antibody sequence. We evaluate the models based on six metrics: Amino Acid Recovery (AAR), Complementarity-determining Amino Acid Recovery (CAAR), TMscore, Local Distance Difference Test (lDDT), Root Mean Square Deviation (RMSD), and DockQ. The detailed experimental settings of this evaluation are provided in the Appendix.

\begin{table}[h!]
\small
\centering
{\renewcommand\baselinestretch{1.5}
\setlength{\tabcolsep}{0.2mm}{
\begin{tabular}{cccccccc}
\toprule
\multirow{2}{*}{Method} & \multicolumn{3}{c}{Generation} & \multicolumn{3}{c}{Docking} \\ 
\cmidrule(lr){2-4} \cmidrule(lr){5-7}
  & AAR $\uparrow$ & TMscore $\uparrow$ & lDDT $\uparrow$ & CAAR $\uparrow$ & RMSD $\downarrow$ & DockQ $\uparrow$ \\ 
\midrule
RosettaAb & 31.91\% & 0.6302 & 0.6107 & 15.29\% & 18.44 & 0.044 \\
HERN & 32.04\% & -  & -  & 17.94\% & 15.47 & 0.056  \\
DiffAb & 24.01\% & 0.6306 & 0.6026 & 18.57\% & 16.60 & 0.039\\
MEAN &37.26\% & 0.6334 & 0.6319 & 23.71\% & 17.32 & 0.047\\
\hline
dyMEAN & 35.97\% & 0.4584 & 0.2657 & 24.11\% & 11.67 & 0.313  \\
\rowcolor[HTML]{E7ECE4} dyAb  & \textbf{37.89\%} & \textbf{0.9264} & \textbf{0.6957} & \textbf{26.14\%} & \textbf{9.86} & \textbf{0.342} \\
\bottomrule
\end{tabular}
} 
\par}
\caption{Results of epitope-binding CDR-H3 design with dynamic antigens on the RAbD benchmark. The first four approaches are multi-stage pipelines, while the last two are end-to-end models.}
\label{tab:cdrh3}
\end{table}

\begin{figure*}
  \centering
  \includegraphics[width=\textwidth]{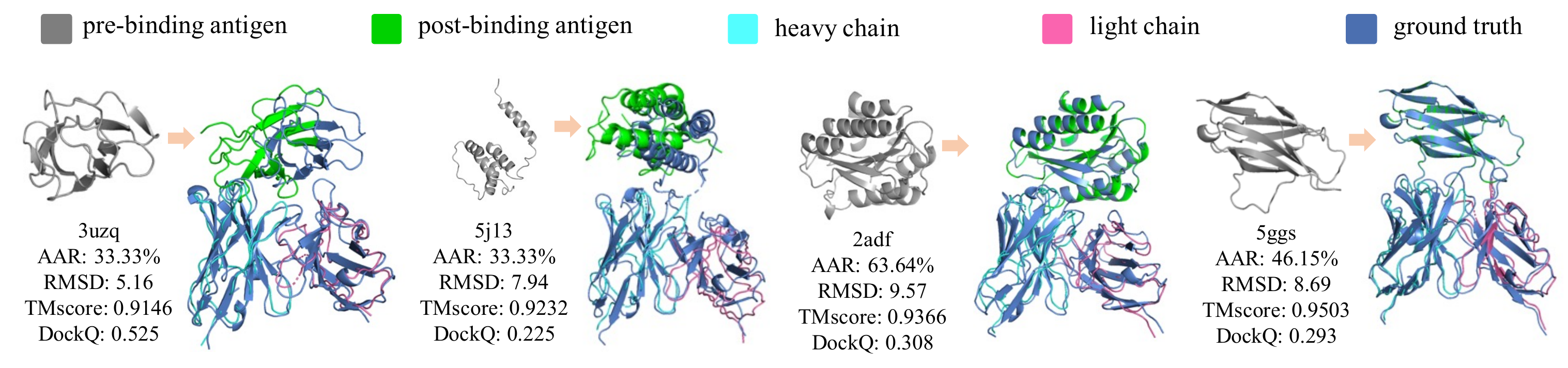}

  \caption{Visualization examples of the generated antibody-antigen complex structures.}

  \label{fig:visualization}
\end{figure*}

We summarize the results in Table~\ref{tab:cdrh3}, showing that dyAb outperforms other methods across various metrics. Compared to dyMEAN, dyAb significantly improves structural metrics like TMscore, lDDT, and RMSD, highlighting its effectiveness in modeling antigen-antibody interactions. Multi-stage methods like RosettaAb, HERN, DiffAb, and MEAN perform worse in DockQ scores because they generate CDR-H3 using pre-binding antigen structures, ignoring conformational changes upon binding. This mismatch leads to lower docking quality. Figure~\ref{fig:visualization} provides examples of dyAb's ability to predict antigen conformational changes and design accurate antibody-antigen structures. More examples are available in the Appendix.


\subsection{Affinity Optimization}

Affinity optimization is a critical task in antibody design, focusing on enhancing the binding affinity of a given antibody-antigen complex. we quantify binding affinity changes using the metric $\Delta\Delta G$ on the SKEMPI dataset, which represents the change in free energy upon binding. Consistent with previous works~\cite{tan2024cross,kong2023end,kong2023conditional}, we employed a $\Delta \Delta G$ predictor~\cite{shan2022deep} and reported the number of average residue changes $\Delta L$ in the optimization process because fewer changes are favored.

We summarize the results in Table~\ref{tab:affinity_optimization}, which presents the outcomes for dyAb and dyMEAN under various constraints on the number of changing residues: 1, 2, 4, 8, and no limit. As we have adapted dyMEAN to take the antigen conformational changes into account, both models show good performance. However, dyAb consistently outperforms dyMEAN across all metrics, achieving superior binding affinity improvements with fewer residue changes. These results suggest that dyAb optimizes antibody-antigen interactions more effectively while minimizing structural alterations.

\begin{table}[ht]
\centering
\setlength{\tabcolsep}{1mm} 
\begin{tabular}{@{}lcccccc@{}}
\toprule
Method & \multicolumn{2}{c}{dyMEAN} & \multicolumn{2}{c}{dyAb} \\
\cmidrule(r){2-3} \cmidrule(r){4-5}
       & $\Delta\Delta G\downarrow$ & $\Delta L \downarrow$ & $\Delta\Delta G\downarrow$ & $\Delta L \downarrow$ \\
\midrule
1      & -9.59 & 0.98 & -13.23 & 0.67 \\
2      & -10.34 & 1.87 & -13.93 & 1.37 \\
4      & -10.77 & 4.79 & -13.88 & 1.60 \\
8      & -10.89 & 6.53 & -14.04 & 3.78 \\
Overall & -11.13 & 6.84 & \textbf{-14.44} & 3.44 \\
\bottomrule
\end{tabular}
\caption{Comparison on binding affinity optimization under different constraints.}
\label{tab:affinity_optimization}
\end{table}




\subsection{Complex Structure Prediction}

The accurate prediction of antibody-antigen complex structures is critical in antibody design. Understanding the precise 3D arrangement of these complexes allows for insights into the binding interactions and mechanisms, which are crucial for developing high-affinity antibodies. We trained our model on the SAbDab dataset and evaluated it on the test set provided by IgFold \cite{ruffolo2021deciphering}. Table~\ref{tab:complex_structure_prediction} presents the results of complex structure prediction. The w/ AF2 in the multi-stage pipeline denotes using AlphaFold2's predicted antigen structure, whereas w/ GT indicates using the ground-truth post-binding antigen structure. It can be seen that dyAb outperforms the other end-to-end baseline, dyMEAN, in all metrics. Furthermore, dyAb achieves comparable results to the multi-stage approaches in TMscore and surpasses them in RMSD and DockQ metrics, highlighting its robustness in predicting the overall structure and docking quality. The relatively lower lDDT score for dyAb may be attributed to the focus on global structural accuracy over local arrangement details.

\begin{table}[h]
\small
\centering
{\renewcommand\baselinestretch{1.2}
\setlength{\tabcolsep}{0.6mm}{
\begin{tabular}{lccccc}
\toprule
\multirow{2}{*}{Method} & \multicolumn{2}{c}{Structure} & \multicolumn{2}{c}{Docking} \\
\cmidrule(lr){2-3} \cmidrule(lr){4-5} & TMscore $\uparrow$ & lDDT $\uparrow$ & RMSD $\downarrow$ & DockQ $\uparrow$ \\
\midrule
IgFold$\rightarrow$HDock, w/ AF2 & 0.6768      & 0.8376        & 17.21       & 0.245         \\
IgFold$\rightarrow$HDock, w/ GT & 0.9502      & 0.8362        & 16.82       & 0.199         \\
IgFold$\rightarrow$HERN, w/ AF2  & 0.8192      & 0.8251        & 13.44        & 0.367          \\
\rowcolor{gray!10} IgFold$\rightarrow$HERN, w/ GT  & \textbf{0.9702}      & \textbf{0.8441}       & \textbf{9.63}        & \textbf{0.429}          \\
\hline
dyMEAN                   & 0.2450      & 0.1564        & 10.39         & 0.391         \\
\rowcolor[HTML]{E7ECE4} dyAb & \textbf{0.9224} & \textbf{0.6871} & \textbf{9.13} & \textbf{0.446} \\
\bottomrule
\end{tabular}
} 
\par}
\caption{Results of complex structure prediction.}
\label{tab:complex_structure_prediction}
\end{table}

\subsection{Ablation Study}
\label{ablation}

We conduct an ablation study and summarize the results in Table~\ref{tab:text_ablation_cdrh3}. Due to the limited space, we leave the detailed analysis in Appendix. Our findings underscore the pivotal role of coarse-grained flow matching in establishing a fundamental conformation. The absence of this alignment significantly degrades both generation and docking performance. Additionally, fine-grained flow matching is crucial for refining the initial conformation, as its removal results in a substantial decline in structural metrics.

\begin{table}[ht]

\small
\centering
{\renewcommand\baselinestretch{1.2}
\setlength{\tabcolsep}{2mm}{
\begin{tabular}{cccccccc}
\toprule
\multirow{2}{*}{Method} & \multicolumn{3}{c}{Generation}\\ 
  & AAR $\uparrow$ & TMscore $\uparrow$ & lDDT $\uparrow$ \\ 
\midrule
dyAb  & 37.89\% & 0.9264 & 0.6957 \\
\hline
dyAb w/o coarse-grained &  13.44\% & 0.1403 & 0.0108 \\
dyAb w/o fine-grained & 37.54\% &	0.5072	& 0.2941   \\
dyAb w/ sampling step=1 & 31.53\% & 0.8932 & 0.6009  \\
dyAb w/ sampling step=50 & 32.80\% & 0.9221 & 0.6793 \\
\bottomrule
\end{tabular}
} 
\par}
\caption{Ablation study on CDR-H3 generation.}
\label{tab:text_ablation_cdrh3}
\end{table}  

Furthermore, we examine the impact of the sampling step size in the flow-matching process. Although the default sampling step is set to 10, we also assess performance using smaller steps of 1 and larger steps of 50. Insufficient sampling with smaller steps and suboptimal performance with larger steps indicate that the sampling size significantly influences performance. Nevertheless, the overall performance across different sampling sizes remains comparable to the default setting, suggesting that while sampling size is crucial, the chosen default size of 10 is effective.

\section{Conclusion and Limitation}

Combining AlphaFold-driven predictions with coarse-grained interface alignment and fine-grained sequence-structure flow matching, dyAb models binding processes with high accuracy. It ensures high-affinity binding and realistic structural evolution, outperforming existing models in dynamic antigen scenarios and streamlining flexible antibody design. However, dyAb's capability still needs to be further examined by wet lab experiments.

 \section{Acknowledgments}

This work was supported by National Science and Technology Major Project (No. 2022ZD0115101), National Natural Science Foundation of China Project (No. 624B2115, No. U21A20427), Project (No. WU2022A009) from the Center of Synthetic Biology and Integrated Bioengineering of Westlake University, Project (No. WU2023C019) from the Westlake University Industries of the Future Research Funding, Tencent AI Lab Rhino-Bird Focused Research Program (Tencent AI Lab RBFR2023007) and an AI\&Health research chair from the Fonds de recherche du Québec en Santé.
 
\bibliography{aaai25}

\begin{thebibliography}{56}
\providecommand{\natexlab}[1]{#1}

\bibitem[{Abramson et~al.(2024)Abramson, Adler, Dunger, Evans, Green, Pritzel, Ronneberger, Willmore, Ballard, Bambrick et~al.}]{abramson2024accurate}
Abramson, J.; Adler, J.; Dunger, J.; Evans, R.; Green, T.; Pritzel, A.; Ronneberger, O.; Willmore, L.; Ballard, A.~J.; Bambrick, J.; et~al. 2024.
\newblock Accurate structure prediction of biomolecular interactions with AlphaFold 3.
\newblock \emph{Nature}, 1--3.

\bibitem[{Adolf-Bryfogle et~al.(2018)Adolf-Bryfogle, Kalyuzhniy, Kubitz, Weitzner, Hu, Adachi, Schief, and Dunbrack~Jr}]{adolf2018rosettaantibodydesign}
Adolf-Bryfogle, J.; Kalyuzhniy, O.; Kubitz, M.; Weitzner, B.~D.; Hu, X.; Adachi, Y.; Schief, W.~R.; and Dunbrack~Jr, R.~L. 2018.
\newblock RosettaAntibodyDesign (RAbD): A general framework for computational antibody design.
\newblock \emph{PLoS computational biology}, 14(4): e1006112.

\bibitem[{Akbar et~al.(2022{\natexlab{a}})Akbar, Bashour, Rawat, Robert, Smorodina, Cotet, Flem-Karlsen, Frank, Mehta, Vu et~al.}]{akbar2022progress}
Akbar, R.; Bashour, H.; Rawat, P.; Robert, P.~A.; Smorodina, E.; Cotet, T.-S.; Flem-Karlsen, K.; Frank, R.; Mehta, B.~B.; Vu, M.~H.; et~al. 2022{\natexlab{a}}.
\newblock Progress and challenges for the machine learning-based design of fit-for-purpose monoclonal antibodies.
\newblock In \emph{MAbs}, volume~14, 2008790. Taylor \& Francis.

\bibitem[{Akbar et~al.(2022{\natexlab{b}})Akbar, Robert, Weber, Widrich, Frank, Pavlovi{\'c}, Scheffer, Chernigovskaya, Snapkov, Slabodkin et~al.}]{akbar2022silico}
Akbar, R.; Robert, P.~A.; Weber, C.~R.; Widrich, M.; Frank, R.; Pavlovi{\'c}, M.; Scheffer, L.; Chernigovskaya, M.; Snapkov, I.; Slabodkin, A.; et~al. 2022{\natexlab{b}}.
\newblock In silico proof of principle of machine learning-based antibody design at unconstrained scale.
\newblock In \emph{MAbs}, volume~14, 2031482. Taylor \& Francis.

\bibitem[{Albergo, Boffi, and Vanden-Eijnden(2023)}]{albergo2023stochastic}
Albergo, M.~S.; Boffi, N.~M.; and Vanden-Eijnden, E. 2023.
\newblock Stochastic interpolants: A unifying framework for flows and diffusions.
\newblock \emph{arXiv preprint arXiv:2303.08797}.

\bibitem[{Alley et~al.(2019)Alley, Khimulya, Biswas, AlQuraishi, and Church}]{alley2019unified}
Alley, E.~C.; Khimulya, G.; Biswas, S.; AlQuraishi, M.; and Church, G.~M. 2019.
\newblock Unified rational protein engineering with sequence-based deep representation learning.
\newblock \emph{Nature methods}, 16(12): 1315--1322.

\bibitem[{Basu et~al.(2019)Basu, Green, Cheng, and Craik}]{basu2019recombinant}
Basu, K.; Green, E.~M.; Cheng, Y.; and Craik, C.~S. 2019.
\newblock Why recombinant antibodies—benefits and applications.
\newblock \emph{Current opinion in biotechnology}, 60: 153--158.

\bibitem[{Das and Baker(2008)}]{das2008macromolecular}
Das, R.; and Baker, D. 2008.
\newblock Macromolecular modeling with rosetta.
\newblock \emph{Annu. Rev. Biochem.}, 77: 363--382.

\bibitem[{Das et~al.(1994)Das, Mavriplis, Saltz, Gupta, and Ponnusamy}]{das1994design}
Das, R.; Mavriplis, D.; Saltz, J.; Gupta, S.; and Ponnusamy, R. 1994.
\newblock Design and implementation of a parallel unstructured Euler solver using software primitives.
\newblock \emph{AIAA journal}, 32(3): 489--496.

\bibitem[{Dauparas et~al.(2022)Dauparas, Anishchenko, Bennett, Bai, Ragotte, Milles, Wicky, Courbet, de~Haas, Bethel et~al.}]{dauparas2022robust}
Dauparas, J.; Anishchenko, I.; Bennett, N.; Bai, H.; Ragotte, R.~J.; Milles, L.~F.; Wicky, B.~I.; Courbet, A.; de~Haas, R.~J.; Bethel, N.; et~al. 2022.
\newblock Robust deep learning--based protein sequence design using ProteinMPNN.
\newblock \emph{Science}, 378(6615): 49--56.

\bibitem[{Del~Alamo et~al.(2022)Del~Alamo, Sala, Mchaourab, and Meiler}]{del2022sampling}
Del~Alamo, D.; Sala, D.; Mchaourab, H.~S.; and Meiler, J. 2022.
\newblock Sampling alternative conformational states of transporters and receptors with AlphaFold2.
\newblock \emph{Elife}, 11: e75751.

\bibitem[{Dunbar et~al.(2014)Dunbar, Krawczyk, Leem, Baker, Fuchs, Georges, Shi, and Deane}]{dunbar2014sabdab}
Dunbar, J.; Krawczyk, K.; Leem, J.; Baker, T.; Fuchs, A.; Georges, G.; Shi, J.; and Deane, C.~M. 2014.
\newblock SAbDab: the structural antibody database.
\newblock \emph{Nucleic acids research}, 42(D1): D1140--D1146.

\bibitem[{Fu and Sun(2022)}]{fu2022antibody}
Fu, T.; and Sun, J. 2022.
\newblock Antibody complementarity determining regions (cdrs) design using constrained energy model.
\newblock In \emph{Proceedings of the 28th ACM SIGKDD Conference on Knowledge Discovery and Data Mining}, 389--399.

\bibitem[{Gao et~al.(2022)Gao, Tan, Li et~al.}]{gao2022alphadesign}
Gao, Z.; Tan, C.; Li, S.; et~al. 2022.
\newblock AlphaDesign: A graph protein design method and benchmark on AlphaFoldDB.
\newblock \emph{arXiv preprint arXiv:2202.01079}.

\bibitem[{Gao, Tan, and Li(2023)}]{gao2023pifold}
Gao, Z.; Tan, C.; and Li, S.~Z. 2023.
\newblock PiFold: Toward effective and efficient protein inverse folding.
\newblock In \emph{ICLR}.

\bibitem[{Han et~al.(2024)Han, Huang, Luo, Han, Shen, Zhang, Zhou, and Chen}]{han2024hemenet}
Han, R.; Huang, W.; Luo, L.; Han, X.; Shen, J.; Zhang, Z.; Zhou, J.; and Chen, T. 2024.
\newblock HeMeNet: Heterogeneous Multichannel Equivariant Network for Protein Multitask Learning.
\newblock \emph{arXiv preprint arXiv:2404.01693}.

\bibitem[{Ho, Jain, and Abbeel(2020)}]{ho2020denoising}
Ho, J.; Jain, A.; and Abbeel, P. 2020.
\newblock Denoising diffusion probabilistic models.
\newblock \emph{NeurIPS}, 33: 6840--6851.

\bibitem[{Hsu et~al.(2022)Hsu, Verkuil, Liu, Lin, Hie, Sercu, Lerer, and Rives}]{hsu2022learning}
Hsu, C.; Verkuil, R.; Liu, J.; Lin, Z.; Hie, B.; Sercu, T.; Lerer, A.; and Rives, A. 2022.
\newblock Learning inverse folding from millions of predicted structures.
\newblock In \emph{ICML}, 8946--8970. PMLR.

\bibitem[{Hu et~al.(2022)Hu, Xia, Zheng, Tan, Huang, Xu, and Li}]{hu2022protein}
Hu, B.; Xia, J.; Zheng, J.; Tan, C.; Huang, Y.; Xu, Y.; and Li, S.~Z. 2022.
\newblock Protein Language Models and Structure Prediction: Connection and Progression.
\newblock arXiv:2211.16742.

\bibitem[{Huber(1992)}]{huber1992robust}
Huber, P.~J. 1992.
\newblock Robust estimation of a location parameter.
\newblock \emph{Breakthroughs in statistics: Methodology and distribution}, 492--518.

\bibitem[{Ingraham et~al.(2019)Ingraham, Garg, Barzilay, and Jaakkola}]{ingraham2019generative}
Ingraham, J.; Garg, V.; Barzilay, R.; and Jaakkola, T. 2019.
\newblock Generative models for graph-based protein design.
\newblock \emph{NeurIPS}, 32.

\bibitem[{Jin, Barzilay, and Jaakkola(2022)}]{jin2022antibody}
Jin, W.; Barzilay, R.; and Jaakkola, T. 2022.
\newblock Antibody-antigen docking and design via hierarchical equivariant refinement.
\newblock \emph{arXiv preprint arXiv:2207.06616}.

\bibitem[{Jin et~al.(2022)Jin, Wohlwend, Barzilay, and Jaakkola}]{jiniterative}
Jin, W.; Wohlwend, J.; Barzilay, R.; and Jaakkola, T.~S. 2022.
\newblock Iterative Refinement Graph Neural Network for Antibody Sequence-Structure Co-design.
\newblock In \emph{ICLR}.

\bibitem[{Jing, Berger, and Jaakkola(2024)}]{jing2024alphafold}
Jing, B.; Berger, B.; and Jaakkola, T. 2024.
\newblock AlphaFold Meets Flow Matching for Generating Protein Ensembles.
\newblock \emph{arXiv preprint arXiv:2402.04845}.

\bibitem[{Jing et~al.(2021)Jing, Eismann, Suriana, Townshend, and Dror}]{jinglearning}
Jing, B.; Eismann, S.; Suriana, P.; Townshend, R. J.~L.; and Dror, R. 2021.
\newblock Learning from Protein Structure with Geometric Vector Perceptrons.
\newblock In \emph{ICLR}.

\bibitem[{Jumper et~al.(2021)Jumper, Evans, Pritzel, Green, Figurnov, Ronneberger, Tunyasuvunakool, Bates, {\v{Z}}{\'\i}dek, Potapenko et~al.}]{jumper2021highly}
Jumper, J.; Evans, R.; Pritzel, A.; Green, T.; Figurnov, M.; Ronneberger, O.; Tunyasuvunakool, K.; Bates, R.; {\v{Z}}{\'\i}dek, A.; Potapenko, A.; et~al. 2021.
\newblock Highly accurate protein structure prediction with AlphaFold.
\newblock \emph{Nature}, 596(7873): 583--589.

\bibitem[{Kabsch(1976)}]{kabsch1976solution}
Kabsch, W. 1976.
\newblock A solution for the best rotation to relate two sets of vectors.
\newblock \emph{Acta Crystallographica Section A: Crystal Physics, Diffraction, Theoretical and General Crystallography}, 32(5): 922--923.

\bibitem[{Kim, Kim, and Park(2024)}]{kim2024anfinsen}
Kim, N.; Kim, M.; and Park, J. 2024.
\newblock Anfinsen Goes Neural: a Graphical Model for Conditional Antibody Design.
\newblock \emph{arXiv preprint arXiv:2402.05982}.

\bibitem[{Kong, Huang, and Liu(2023{\natexlab{a}})}]{kong2023conditional}
Kong, X.; Huang, W.; and Liu, Y. 2023{\natexlab{a}}.
\newblock Conditional Antibody Design as 3D Equivariant Graph Translation.
\newblock In \emph{The Eleventh ICLR}.

\bibitem[{Kong, Huang, and Liu(2023{\natexlab{b}})}]{kong2023end}
Kong, X.; Huang, W.; and Liu, Y. 2023{\natexlab{b}}.
\newblock End-to-End Full-Atom Antibody Design.
\newblock \emph{arXiv preprint arXiv:2302.00203}.

\bibitem[{Kong, Huang, and Liu(2024)}]{kong2024full}
Kong, X.; Huang, W.; and Liu, Y. 2024.
\newblock Full-Atom Peptide Design with Geometric Latent Diffusion.
\newblock \emph{arXiv preprint arXiv:2402.13555}.

\bibitem[{Kuroda et~al.(2012)Kuroda, Shirai, Jacobson, and Nakamura}]{kuroda2012computer}
Kuroda, D.; Shirai, H.; Jacobson, M.~P.; and Nakamura, H. 2012.
\newblock Computer-aided antibody design.
\newblock \emph{Protein engineering, design \& selection}, 25(10): 507--522.

\bibitem[{Lefranc et~al.(2003)Lefranc, Pommi{\'e}, Ruiz, Giudicelli, Foulquier, Truong, Thouvenin-Contet, and Lefranc}]{lefranc2003imgt}
Lefranc, M.-P.; Pommi{\'e}, C.; Ruiz, M.; Giudicelli, V.; Foulquier, E.; Truong, L.; Thouvenin-Contet, V.; and Lefranc, G. 2003.
\newblock IMGT unique numbering for immunoglobulin and T cell receptor variable domains and Ig superfamily V-like domains.
\newblock \emph{Developmental \& Comparative Immunology}, 27(1): 55--77.

\bibitem[{Lin et~al.(2024)Lin, Zhang, Zhao, Wu, Jiang, Liu, Huang, and Li}]{lin2024ppflow}
Lin, H.; Zhang, O.; Zhao, H.; Wu, L.; Jiang, D.; Liu, Z.; Huang, Y.; and Li, S.~Z. 2024.
\newblock PPFlow: Target-aware Peptide Design with Torsional Flow Matching.
\newblock \emph{bioRxiv}, 2024--03.

\bibitem[{Lipman et~al.(2022)Lipman, Chen, Ben-Hamu, Nickel, and Le}]{lipman2022flow}
Lipman, Y.; Chen, R.~T.; Ben-Hamu, H.; Nickel, M.; and Le, M. 2022.
\newblock Flow Matching for Generative Modeling.
\newblock In \emph{The Eleventh ICLR}.

\bibitem[{Liu, Gong et~al.(2022)}]{liu2022flow}
Liu, X.; Gong, C.; et~al. 2022.
\newblock Flow Straight and Fast: Learning to Generate and Transfer Data with Rectified Flow.
\newblock In \emph{The Eleventh ICLR}.

\bibitem[{Luo et~al.(2022)Luo, Su, Peng, Wang, Peng, and Ma}]{luoantigen}
Luo, S.; Su, Y.; Peng, X.; Wang, S.; Peng, J.; and Ma, J. 2022.
\newblock Antigen-Specific Antibody Design and Optimization with Diffusion-Based Generative Models for Protein Structures.
\newblock In \emph{NeurIPS}.

\bibitem[{Maynard and Georgiou(2000)}]{maynard2000antibody}
Maynard, J.; and Georgiou, G. 2000.
\newblock Antibody engineering.
\newblock \emph{Annual review of biomedical engineering}, 2(1): 339--376.

\bibitem[{Pantazes and Maranas(2010)}]{pantazes2010optcdr}
Pantazes, R.; and Maranas, C.~D. 2010.
\newblock OptCDR: a general computational method for the design of antibody complementarity determining regions for targeted epitope binding.
\newblock \emph{Protein Engineering, Design \& Selection}, 23(11): 849--858.

\bibitem[{Raybould et~al.(2019)Raybould, Marks, Krawczyk, Taddese, Nowak, Lewis, Bujotzek, Shi, and Deane}]{raybould2019five}
Raybould, M.~I.; Marks, C.; Krawczyk, K.; Taddese, B.; Nowak, J.; Lewis, A.~P.; Bujotzek, A.; Shi, J.; and Deane, C.~M. 2019.
\newblock Five computational developability guidelines for therapeutic antibody profiling.
\newblock \emph{Proceedings of the National Academy of Sciences}, 116(10): 4025--4030.

\bibitem[{Ruffolo, Gray, and Sulam(2021)}]{ruffolo2021deciphering}
Ruffolo, J.~A.; Gray, J.~J.; and Sulam, J. 2021.
\newblock Deciphering antibody affinity maturation with language models and weakly supervised learning.
\newblock \emph{arXiv preprint arXiv:2112.07782}.

\bibitem[{Saka et~al.(2021)Saka, Kakuzaki, Metsugi, Kashiwagi, Yoshida, Wada, Tsunoda, and Teramoto}]{saka2021antibody}
Saka, K.; Kakuzaki, T.; Metsugi, S.; Kashiwagi, D.; Yoshida, K.; Wada, M.; Tsunoda, H.; and Teramoto, R. 2021.
\newblock Antibody design using LSTM based deep generative model from phage display library for affinity maturation.
\newblock \emph{Scientific reports}, 11(1): 1--13.

\bibitem[{Shan et~al.(2022)Shan, Luo, Yang, Hong, Su, Ding, Fu, Li, Chen, Ma et~al.}]{shan2022deep}
Shan, S.; Luo, S.; Yang, Z.; Hong, J.; Su, Y.; Ding, F.; Fu, L.; Li, C.; Chen, P.; Ma, J.; et~al. 2022.
\newblock Deep learning guided optimization of human antibody against SARS-CoV-2 variants with broad neutralization.
\newblock \emph{Proceedings of the National Academy of Sciences}, 119(11): e2122954119.

\bibitem[{Shi et~al.(2022)Shi, Wang, Lu, Zhong, and Tang}]{shi2022protein}
Shi, C.; Wang, C.; Lu, J.; Zhong, B.; and Tang, J. 2022.
\newblock Protein sequence and structure co-design with equivariant translation.
\newblock \emph{arXiv preprint arXiv:2210.08761}.

\bibitem[{Shin et~al.(2021)Shin, Riesselman, Kollasch, McMahon, Simon, Sander, Manglik, Kruse, and Marks}]{shin2021protein}
Shin, J.-E.; Riesselman, A.~J.; Kollasch, A.~W.; McMahon, C.; Simon, E.; Sander, C.; Manglik, A.; Kruse, A.~C.; and Marks, D.~S. 2021.
\newblock Protein design and variant prediction using autoregressive generative models.
\newblock \emph{Nature communications}, 12(1): 2403.

\bibitem[{Song et~al.(2021)Song, Sohl-Dickstein, Kingma, Kumar, Ermon, and Poole}]{song2021score}
Song, Y.; Sohl-Dickstein, J.; Kingma, D.~P.; Kumar, A.; Ermon, S.; and Poole, B. 2021.
\newblock Score-Based Generative Modeling through Stochastic Differential Equations.
\newblock In \emph{ICLR}.

\bibitem[{St{\"a}rk et~al.(2023)St{\"a}rk, Jing, Barzilay, and Jaakkola}]{stark2023harmonic}
St{\"a}rk, H.; Jing, B.; Barzilay, R.; and Jaakkola, T. 2023.
\newblock Harmonic Self-Conditioned Flow Matching for Multi-Ligand Docking and Binding Site Design.
\newblock \emph{arXiv preprint arXiv:2310.05764}.

\bibitem[{Tan et~al.(2024)Tan, Gao, Wu, Xia, Zheng, Yang, Liu, Hu, and Li}]{tan2024cross}
Tan, C.; Gao, Z.; Wu, L.; Xia, J.; Zheng, J.; Yang, X.; Liu, Y.; Hu, B.; and Li, S.~Z. 2024.
\newblock Cross-Gate MLP with Protein Complex Invariant Embedding Is a One-Shot Antibody Designer.
\newblock In \emph{Proceedings of the AAAI Conference on Artificial Intelligence}, volume~38, 15222--15230.

\bibitem[{Tan et~al.(2023)Tan, Gao, Xia, Hu, and Li}]{tan2022generative}
Tan, C.; Gao, Z.; Xia, J.; Hu, B.; and Li, S.~Z. 2023.
\newblock Generative de novo protein design with global context.
\newblock In \emph{ICASSP 2023-2023 IEEE International Conference on Acoustics, Speech and Signal Processing (ICASSP)}. IEEE.

\bibitem[{Tiller and Tessier(2015)}]{tiller2015advances}
Tiller, K.~E.; and Tessier, P.~M. 2015.
\newblock Advances in antibody design.
\newblock \emph{Annual review of biomedical engineering}, 17: 191--216.

\bibitem[{Varadi et~al.(2022)Varadi, Anyango, Deshpande, Nair, Natassia, Yordanova, Yuan, Stroe, Wood, Laydon et~al.}]{varadi2022alphafold}
Varadi, M.; Anyango, S.; Deshpande, M.; Nair, S.; Natassia, C.; Yordanova, G.; Yuan, D.; Stroe, O.; Wood, G.; Laydon, A.; et~al. 2022.
\newblock AlphaFold Protein Structure Database: massively expanding the structural coverage of protein-sequence space with high-accuracy models.
\newblock \emph{Nucleic acids research}, 50(D1): D439--D444.

\bibitem[{Wang et~al.(2021)Wang, Xue, Cao, Yu, Lane, and Zhao}]{wang2021directed}
Wang, Y.; Xue, P.; Cao, M.; Yu, T.; Lane, S.~T.; and Zhao, H. 2021.
\newblock Directed evolution: methodologies and applications.
\newblock \emph{Chemical reviews}, 121(20): 12384--12444.

\bibitem[{Warszawski et~al.(2019)Warszawski, Borenstein~Katz, Lipsh, Khmelnitsky, Ben~Nissan, Javitt, Dym, Unger, Knop, Albeck et~al.}]{warszawski2019optimizing}
Warszawski, S.; Borenstein~Katz, A.; Lipsh, R.; Khmelnitsky, L.; Ben~Nissan, G.; Javitt, G.; Dym, O.; Unger, T.; Knop, O.; Albeck, S.; et~al. 2019.
\newblock Optimizing antibody affinity and stability by the automated design of the variable light-heavy chain interfaces.
\newblock \emph{PLoS computational biology}, 15(8): e1007207.

\bibitem[{Wayment-Steele et~al.(2024)Wayment-Steele, Ojoawo, Otten, Apitz, Pitsawong, H{\"o}mberger, Ovchinnikov, Colwell, and Kern}]{wayment2024predicting}
Wayment-Steele, H.~K.; Ojoawo, A.; Otten, R.; Apitz, J.~M.; Pitsawong, W.; H{\"o}mberger, M.; Ovchinnikov, S.; Colwell, L.; and Kern, D. 2024.
\newblock Predicting multiple conformations via sequence clustering and AlphaFold2.
\newblock \emph{Nature}, 625(7996): 832--839.

\bibitem[{Wu et~al.(2022)Wu, Wu, Jiang, Liu, and Zhao}]{wu2022tfold}
Wu, J.; Wu, F.; Jiang, B.; Liu, W.; and Zhao, P. 2022.
\newblock tFold-Ab: fast and accurate antibody structure prediction without sequence homologs.
\newblock \emph{Biorxiv}, 2022--11.

\bibitem[{Yan et~al.(2017)Yan, Zhang, Zhou, Li, and Huang}]{yan2017hdock}
Yan, Y.; Zhang, D.; Zhou, P.; Li, B.; and Huang, S.-Y. 2017.
\newblock HDOCK: a web server for protein--protein and protein--DNA/RNA docking based on a hybrid strategy.
\newblock \emph{Nucleic acids research}, 45(W1): W365--W373.

\end{thebibliography}

\end{document}